\pdfoutput=1

\documentclass[fleqn,usenatbib]{mnras}

\usepackage{newtxtext,newtxmath}

\usepackage[T1]{fontenc}
\usepackage{ae,aecompl}

\usepackage[dvipsnames]{xcolor}

\usepackage{graphicx}	
\usepackage{amsmath}	
\usepackage{listings}
\usepackage{soul}

\newcommand{\fink}{{\sc Fink}}

\title[\fink, a new generation broker for the LSST community]{\fink, a new generation of broker for the LSST community}

\author[M{\"o}ller, Peloton, Ishida et al.]{Anais M{\"o}ller$^{1}$\thanks{E-mail:anais.moller@clermont.in2p3.fr},
Julien Peloton$^{2}$\thanks{E-mail:peloton@lal.in2p3.fr},
Emille E. O. Ishida$^{1}$\thanks{E-mail:emille.ishida@clermont.in2p3.fr}, 
\newauthor
Chris Arnault$^{2}$,
Etienne Bachelet $^{3}$,
Tristan Blaineau$^{2}$,
Dominique Boutigny$^{4}$,
\newauthor
Abhishek Chauhan$^{5}$,
Emmanuel Gangler$^{1}$,
Fabio Hernandez$^{6}$,
Julius Hrivnac$^{2}$,
\newauthor
Marco Leoni$^{2,7}$,
Nicolas Leroy$^{2}$,
Marc Moniez$^{2}$,
Sacha Pateyron$^{2}$,
Adrien Ramparison$^{2}$,
\newauthor
Damien Turpin$^{8}$,
R{\'e}za Ansari$^{2}$,
Tarek Allam Jr.$^{9,10}$,
Armelle Bajat $^{11}$,
Biswajit Biswas$^{1,12}$,
\newauthor
Alexandre Boucaud$^{13}$,
Johan Bregeon$^{14}$,
Jean-Eric Campagne$^{2}$, 
Johann Cohen-Tanugi$^{15,1}$,
\newauthor
Alexis Coleiro$^{13}$,
Damien Dornic$^{16}$,
Dominique Fouchez$^{16}$,
Olivier Godet$^{18}$,
Philippe Gris$^{1}$,
\newauthor
Sergey Karpov $^{11}$,
Ada Nebot Gomez-Moran $^{18}$,
J{\'e}r{\'e}my Neveu$^{2}$,
\newauthor
Stephane Plaszczynski$^{2}$,
Volodymyr Savchenko$^{19}$,
Natalie Webb$^{17}$\\
$^{1}$LPC, Universit{\'e} Clermont Auvergne, CNRS/IN2P3, F-63000 Clermont-Ferrand, France.\\
$^{2}$Universit{\'e} Paris-Saclay, CNRS/IN2P3, IJCLab, Orsay, France\\
$^{3}$Las Cumbres Observatory, 6740 Cortona Drive, suite 102, Goleta, CA 93117, USA.\\
$^{4}$Universit{\'e} Grenoble-Alpes, Universit{\'e} Savoie Mont Blanc, CNRS/IN2P3 Laboratoire d'Annecy-le-Vieux de Physique des Particules, France\\
$^{5}$Department of Ocean Engineering and Naval Architecture, Indian Institute of Technology Kharagpur, West Bengal, 721302, India\\
$^{6}$CNRS, CC-IN2P3, 21 avenue Pierre de Coubertin, CS70202, 69627 Villeurbanne cedex, France. \\
$^{7}$Direction des Syst{\`e}mes d'Information, Universit{\'e} Paris Sud, 91405 Orsay Cedex, France \\
$^{8}$Universit\'e Paris-Saclay, CNRS, CEA, D\'epartement d'Astrophysique, Instrumentation et Mod\'elisation de Paris-Saclay, Gif-sur-Yvette, France\\
$^{9}$Centre for Data Intensive Science, Department of Physics and Astronomy, University College London, London WC1E 6BT\\
$^{10}$Mullard Space Science Laboratory (MSSL), Department of Space and Climate Physics, University College London, Surrey RH5 6NT, UK.\\
$^{11}$CEICO, Institute of Physics, Czech Academy of Sciences, Prague, Czech Republic. \\
$^{12}$Department of Physics, Birla Institute of Technology and Science, Pilani, Pilani Campus, Rajasthan, 333031, India.\\
$^{13}$Universit{\'e} de Paris, CNRS, AstroParticule et Cosmologie, F-75013, Paris, France\\
$^{14}$CNRS-IN2P3, Laboratoire de Physique Subatomique et de Cosmologie (LPSC), Grenoble \\
$^{15}$LUPM, Universit{\'e} de Montpellier et CNRS, 34095 Montpellier Cedex 05, France.\\
$^{16}$Aix Marseille Universit{\'e}, CNRS/IN2P3, CPPM, Marseille, France\\
$^{17}$IRAP UPS/CNRS/CNES, 9 avenue du colonel Roche 3028 Toulouse Cedex 04, France.\\
$^{18}$Observatoire Astronomique de Strasbourg, Universit{\'e} de Strasbourg, CNRS UMR 7550, 11 rue de l'Universit{\'e} , 67000 Strasbourg, France.\\
$^{19}$ISDC, Department of Astronomy, University of Geneva, Chemin d'Ecogia, 16 CH-1290 Versoix, Switzerland
}

\pubyear{2020}

\begin{document}
\label{firstpage}
\pagerange{\pageref{firstpage}--\pageref{lastpage}}
\maketitle

\begin{abstract}
\fink\ is a broker designed to enable science with large time-domain alert streams such as the one from the upcoming Vera C. Rubin Observatory Legacy Survey of Space and Time (LSST). It exhibits traditional astronomy broker features such as automatised ingestion, annotation, selection and redistribution of promising alerts for transient science. It is also designed to go beyond traditional broker features by providing real-time transient classification which is continuously improved by using state-of-the-art Deep Learning and Adaptive Learning techniques. These evolving added values will enable more accurate scientific output from LSST photometric data for diverse science cases while also leading to a higher incidence of new discoveries which shall accompany the evolution of the survey. In this paper we introduce \fink, its science motivation, architecture and current status including first science verification cases using the Zwicky Transient Facility alert stream.
\end{abstract}

\begin{keywords}
surveys -- methods: data analysis -- software: data analysis  -- transients: gamma-ray bursts -- gravitational lensing: micro -- transients: supernovae
\vspace{-1.5\baselineskip}  
\end{keywords}


\section*{Introduction}
The Vera C. Rubin Observatory Legacy Survey of Space and Time aims to survey the southern sky deeper and faster than any wide-field survey to date. During its 10 years duration, LSST will enable the discovery of an unprecedented large number of astrophysical transients opening a new era of optical big data in astronomy. Varying sources will be identified using a difference imaging analysis pipeline and transient candidates reported within 60 seconds via a public alert stream. This system is expected to produce $\approx 10$ million alerts per night \citep{ldm612}.

To harness the full power of LSST, promising candidates must be identified within this alert stream. This task includes many challenges: first, the stream will contain diverse astrophysical phenomena which are challenging to disentangle. Second, identification of promising candidates must be done in a timely manner due to their transient nature, with time frames spanning from seconds to months. Third, the alert rate forecast for LSST will be at least an order of magnitude larger than current surveys and it will trigger on typically fainter objects, making it difficult for currently available systems to operate efficiently.

\fink \footnote{\fink\ is not an acronym. Email: contact@fink-broker.org} is a community broker specifically developed to address these challenges. It has been conceived and built within a large and interdisciplinary scientific community as an answer to the call for Letters of Intent for Community Alert Brokers \citep{ldm682}, advertised by the Rubin Observatory LSST data management team in 2019\footnote{The \fink\ LOI can be found at this \href{https://owncloud.lal.in2p3.fr/index.php/s/XdQnCWcvcjbQ6Vr}{URL}}. Based on established technologies for fast and efficient analysis of big data, it provides traditional broker features such as catalogue and survey cross-matches, but also uses state-of-the-art machine learning techniques to generate classification scores for a variety of time-domain phenomena. In addition, it is designed to adapt with the evolution of the available information, delivering increasingly  more accurate added values throughout the duration of the survey.

\fink\ comes to join a few other brokers currently operating on other experiments, such as the Zwicky Transient Facility \citep[ZTF, ][]{Bellm_2018} or the High cadence Transient Survey \citep[HiTS, ][]{2016ApJ...832..155F}. Among these are ALeRCE \citep{Forster:2020}, Ampel \citep{ampel}, Antares \citep{antares}, Lasair \citep{Smith_2019}, MARS\footnote{\url{https://mars.lco.global/}} and SkyPortal \citep{skyportal2019}. 
ZTF has the particularity to use an alert system design that is very close to the one envisioned by LSST \citep{DMTN-093}, hence allowing to prototype and test systems with the relevant features and communication protocols prior to the start of LSST operations. 

In this paper we will introduce the \fink\ broker. First, we will showcase the different science cases used for its original development in Section~\ref{section:science}. We then summarise the design choices 
motivated by these science cases and the technological challenges of the LSST alert stream in Section~\ref{section:requirements}. We outline the architecture and current implementation in Section~\ref{section:implementation}, present the first science verification tasks in Section~\ref{section:sciecenverification} and summarise the project and first results in Section~\ref{section:summary}.

\section{Science drivers} \label{section:science}

The goal of \fink\ is to enable discovery in many areas of time-domain astronomy. It has been designed to be flexible and can be used for a variety of science cases, from stellar microlensing, to extra-galactic transients and cosmology. In the following, we present the initial science cases which inspired the design and conception of this project. We are open to contributions in these science cases as well as new contributions that are not listed here for both science drivers and science modules\footnote{Details on how to propose science modules for these and new science cases can be found at \url{https://fink-broker.org/joining.html}} (see Section~\ref{section:sciencemodules}).

\subsection{Multi-messenger astronomy}
The study of the transient sky has entered to a new era, with instruments being able to detect sources in the whole electromagnetic band from radio to TeV photons but also through high energy neutrinos, very high energy cosmic rays and gravitational waves. The transients detected by LSST will also benefit from current space and ground instrumentation like KM3NeT neutrino detectors \citep{Adrian-Martinez:2016fdl}, the extension of LIGO/Virgo and KAGRA for gravitational waves  \citep{Abbott:2018}, the Square Kilometre Array telescopes \citep[SKA, radio bands;][]{SKA:2009}, the Chinese-French Space Variable Objects Monitor multi-wavelength observatory \citep[SVOM, gamma and X rays, and optical ground segment;][]{SVOM:2016}, the Cherenkov Telescope Array \citep[CTA, gamma rays;][]{CTA:2011} and many others.

To fully exploit this abundance of data, it is paramount to connect different surveys together and promote coordination of resources. In the following, we present some of the challenges faced by specific transient phenomena.

\subsubsection*{Gamma-ray bursts}

The study of Gamma-ray Bursts (GRBs) has seen major breakthroughs in the past 2 years with multi-messenger observations of associated events such as GW170817 \citep{Abbott:2017} the H.E.S.S. detection of GRB180720B afterglow \citep{HESS_ATel:2019} and multi-wavelength observations of GRB 190114C detected early by MAGIC and then observed in the optical \citep{MAGIC:2019}.

The main bottleneck in these multi-messenger multi-wavelength observations of GRBs, is quick and precise localisation of the counterpart, including a good estimate of the redshift that can only be achieved through spectroscopic follow-up. In the LSST decade, following the {\it Swift} mission legacy, the SVOM mission will aim at detecting and identifying GRB optical counterparts within hours of the GRB trigger time to propose fast follow-up target observations to spectroscopic facilities. 

Our broker aims to enable {\it Swift} localisation of an optical counterpart using the LSST alert stream. For this, the broker is designed to cross-match GRB alerts from large field-of-view gamma observatories to the LSST optical transient alert stream and provide counterpart candidates within minutes of the optical observation. 

For these science cases, it is crucial to efficiently combine observations from multi-messenger and multi-wavelength surveys. \fink\ will be able to combine the early and mid-term follow-up observations performed by the SVOM instruments with the photometric measurements provided by LSST to characterise the events detected on-board the SVOM spacecraft. LSST photometry could be useful to search for supernovae or kilonovae for nearby GRBs and put constraints in the jet physics.

Furthermore, our broker also aims at identifying the up to 100 orphan GRB optical afterglows expected in LSST per year \citep{Ghirlanda:2015}. Beside the direct scientific return that such detection and identification would provide, Fink will be able to perform cross-matches with relevant external multi-messenger data sets.

\subsubsection*{Gravitational Waves (GW)}
The discovery of a gravitational wave signal from a binary neutron star system (GW170817) in coincidence with a short gamma-ray burst (GRB170817A) and followed by an optical kilonova (AT 2017gfo) showed the importance of multi-messenger astronomy by combining all the information available to study this transient event \citep{Abbott:2017}.

Thanks to its deep large field of view, LSST will be able to catch the very first glimpse of electromagnetic counterparts from gravitational waves (if they are above saturation levels) imposing further constraints on their time evolution models. Furthermore, building a large sample of electromagnetic counterparts of GW events is also important for performing fundamental physics and cosmology tests (e.g. GW velocity and measurement of $H_0$). For such studies, it is crucial that candidates compatible with the sky region provided by the GW network be quickly identified in optical data so follow-up can be triggered. Our broker will select alerts within the GW detection region and provide preliminary classifications, considering kilonovae and other potential transient events. These subsets of alerts will be automatically communicated within minutes of observation to guide optimal distribution of follow-up efforts.

Furthermore, the broker can identify potential kilonovae in the optical and communicate them to GW experiments, thus allowing searches of low signal-to-noise gravitational waves which can be found by re-analysing GW data.

\subsubsection*{High-energy neutrinos}
The discovery of a high-energy cosmic diffuse neutrino flux by IceCube~\citep{Aartsen:2014gkd} and ANTARES~\citep{Fusco:2019tok} and the first evidences of the point-like sources such as the coincidence between the high-energy neutrino IC170922 and a flaring gamma-ray blazar TXS 0506+056~\citep{IceCube:2018cha, IceCube:2018dnn} and the neutrino IC191001A with a tidal disruption event AT2019dsg~\citep{Stein:2020xhk} have initiated a new era in this field. In both associations, some tiny time variations in the optical bands have also been identified.  Since recent years, IceCube \citep[and later KM3NeT;][]{Adrian-Martinez:2016fdl} is sending alerts in the cascade channel (interaction of electron and tau neutrinos and neutral current muon neutrino) with a precision of 5-15 degrees radius.

Thanks to its deep large field of view, its precise photometry and its high cadence, LSST will be able to catch the optical transients associated with all flavour neutrinos and therefore to provide precise coordinates crucial for further follow-up necessary for a precise characterisation of the nature of the source.

For the neutrino application, our broker will select and identify optical transient candidates within the neutrino detection region and provide their preliminary classifications. These subset of alerts will be automatically communicated within minutes of observation to guide optimal distribution of follow-up efforts.

\subsubsection*{Tidal disruption events}

LSST will also lend itself very well to discovering tidal disruption events (TDEs). These occur when a star in a galaxy wanders too close to the central massive black hole. The star disrupts when the tidal forces exceed the self-gravity of the star and a previously undetected central super-massive black hole will become extremely bright, allowing it to be studied. Rates of tidal disruptions are estimated to be 1.7$\pm^{\scriptscriptstyle +2.85}_{\scriptscriptstyle -1.27}$ $\times$ 10$^{-4}$ gal$^{-1}$ yr$^{-1}$ \citep[90\% confidence][]{Hung_2018}. To date, only about 90 TDEs have been detected\footnote{\url{https://tde.space/}}, where about half are detected in the X-rays and most of the others in the optical/infra-red bands. While the X-ray emission can constrain the mass of the black hole, timely spectroscopic follow-up may put limits on the black hole spin \citep[e.g.][]{Karas_2014}. As the tidal radius of the black hole must be outside of the Schwarzschild radius to observe the tidal disruption event, only the lower mass black holes ($<$ 10$^8$ M$_\odot$) are detected \citep{Miller:2004}. This will allow to detect and study the extremely rare intermediate mass black holes (IMBH, $\sim$10$^{2-5}$ M$_\odot$). TDEs studies will also be key to test the scaling to the lower BH mass regime and shed light on the evolution history of dwarf galaxies in the local Universe.

Thousands of TDEs per year will be discovered with LSST. They will help us to determine the demographics of massive black holes as well as finding intermediate mass black holes (there should be many detected per year), which are thought to be the building blocks of super-massive black holes. The challenge our broker will aim to tackle is to identify tidal disruption events in real time for rapid follow-up observations. For this, our broker will incorporate methods to distinguish TDEs from other long transient events (e.g. X-ray binary or cataclysmic variable outbursts or supernova outbursts) using LSST, X-ray and other multi-wavelength data. 

\subsection{Microlensing}
Microlensing occurs when a foreground massive object (lens) passes directly between the observer and a luminous background source. The gravitational potential of the lens deflects the light from the source with a characteristic radius, causing the source to brighten and fade as they move into and out of alignment, with a timescale that is proportional to the square root of the lens mass (years for black holes, days to weeks for planets and stars).

It is expected that LSST will detect more than 5,000 microlensing events during its lifetime \citep{Sajadian2019}, enabling the detection of exoplanets and isolated stellar mass black holes in the Milky Way. 

LSST alone will not provide enough information to study short time-scale distortions in microlensing events. However, potential candidates can be identified using the LSST alert stream. \fink\ will be instrumental in the identification of potential candidates and the coordination of quick follow-up through specific networks like MicroFUN\footnote{\url{http://www.astronomy.ohio-state.edu/~microfun/}} and RoboNET team \footnote{\url{https://robonet.lco.global/}}, or astrometric and spectroscopic follow-up observations (through spatial telescopes and large spectrometers).

\subsection{Supernovae}
Supernovae are stellar explosions of interest for astrophysical and cosmology studies. They are visible during week time-scales in multiple wavelengths and potentially through other messengers such as neutrinos, gravitational waves and gamma rays. LSST is expected to discover up to $10^6$ supernovae during its 10 year survey \citep{LSST:2009}, orders of magnitude greater than currently available data sets.

Given such data volume, the first challenge will be to disentangle different types of supernovae using the alert streams. \fink\ will deliver classification, 
after only a few observations, in order to select promising candidates for further analysis and follow-up coordination \citep[e.g.][]{Moller:2019}. This early identification is crucial to allow optimal distribution of follow-up efforts for further SNe studies, including spectroscopic typing of supernovae, as well as to improve training sets for photometric classifiers \citep{Ishida:2019}. 

For cosmology, the main challenge for any broker will be to characterise its selection function during the full 10 year survey. In other words it is necessary to reconstruct a posteriori the state of the broker at any given time in order to properly evaluate bias corrections. Determining this selection function is the dominant uncertainty on the bias corrections which is one of the top systematics in current cosmology constraints \citep{Betoule:2014,DES:2019}. \fink\ aims to keep a record of the state of the broker thus enabling efficient and traceable selection functions.
Moreover, the broker is being developed in close contact with the LSST Dark Energy Science Collaboration (DESC) Alerts topical team in order to ensure that it fulfils the requirements for LSST dark energy supernova searches. These efforts aim to fully exploit the availability of LSST data at IN2P3 Computing Centre in Lyon (CC-IN2P3) (see more details in Section \ref{section:requirementsanddesign}), which gives \fink\ an opportunity to cross-match transient candidates with LSST data, thus providing extra added values which can be made available to DESC members and boost scientific outcomes of dark energy studies.

\subsection{Anomaly Detection}  
\label{section:anomalies}

Perhaps one of the most anticipated outcomes of the LSST era is the identification of unknown astrophysical sources. As an example, potentially cataclysmic events leading to new mechanisms of particle acceleration or electromagnetic counterparts of GW can provide important insights into their progenitor systems. However, given the volume and complexity of the data which will accompany them, it is unlikely that such sources will be serendipitously identified. The use of automated anomaly detection (AD) algorithms might help mitigate some of these issues. However, traditional AD techniques are known to deliver a high incidence of false positives \citep[see e.g.,][]{Pruzhinskaya:2019}. Taking into account that, in astronomy, further scrutiny of an anomaly candidate is only possible through expensive spectroscopic follow-up, this translates into a significant fraction of resources spent in non-interesting targets. This caveat can be overcome by carefully designing a recommendation system able to identify and flag candidates following a custom anomaly definition. This is the goal of active anomaly detection techniques \citep{Das2017}. 
In order to fully exploit the potential of LSST for new discoveries, \fink\ is designed to have an anomaly detection module, based on state-of-the-art adaptive machine learning techniques which will be tailored to optimise the use of domain knowledge \citep{Ishida_ALanomaly:2019}. This will allow the broker to deliver increasingly more accurate anomaly scores throughout the first years of the survey.

\begin{figure*}
    \centering
    \includegraphics[width=\textwidth]{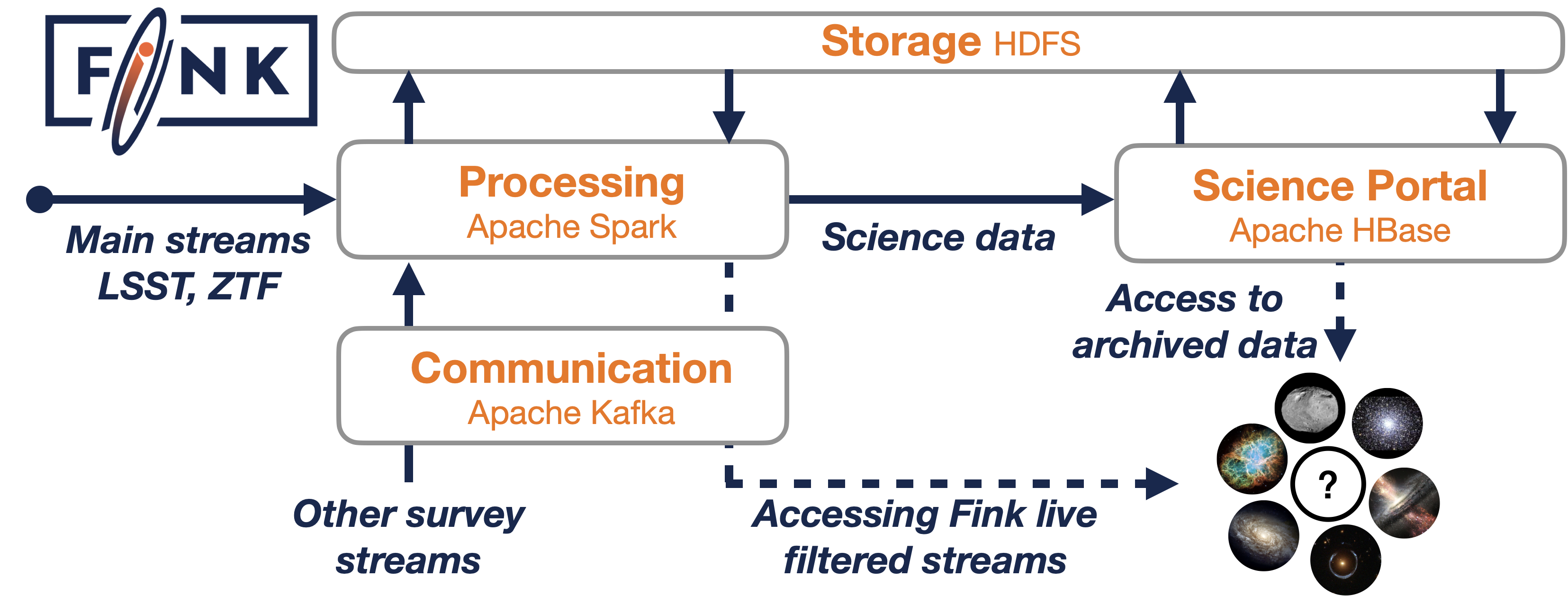}
    \caption{Architecture of \fink. Each box is a cluster of machines deployed on a cloud. The main streams of alerts for \fink\ (ZTF, and LSST) are collected and processed inside the \texttt{Processing} cluster running Apache Spark (see Sec. \ref{section:data-ingestion}, \ref{section:sciencemodules}). At the end of the processing, a series of filter divides the stream into substreams based on user needs, and data is sent to subscribers via the \texttt{Communication} cluster running Apache Kafka (see Sec. \ref{section:filteredstreams}). At the end of the night, all processed data is aggregated and pushed into the \texttt{Science Portal}, based on Apache HBase, where users can connect via a web browser and explore all processed \fink\ data (see Sec. \ref{sec:science-portal}). Alert data and added-values are stored at various stages on the Hadoop Distributed File System (HDFS). Other survey data streams (such as alert data from LIGO/Virgo, Fermi or {\it Swift}) are collected by the \texttt{Communication} cluster and sent to the \texttt{Processing} cluster to be used to annotate the main stream of alerts. See text for more information on each component.}
    \label{figure:architecture}
\end{figure*}

\section{Requirements \& design}\label{section:requirementsanddesign}

\fink\ is designed to maximise scientific exploitation of the LSST alert stream. We introduce its science and technology requirements in Section~\ref{section:requirements}. Chosen big data technologies are introduced in  Section~\ref{section:bigdata}. We make a strong emphasis on interoperability with existing frameworks for catalogue query, cross-matching and communication between surveys and follow-up facilities, as we show in Section~\ref{section:interoperability}. In Section~\ref{section:modularity} we focus on modularity and versioning which enable the system to adapt to the evolution of tools and communities over a decade, as well ensuring traceability of its delivered data and added values.

\subsection{Requirements} \label{section:requirements}

\fink\ fulfils traditional broker tasks, for the above science cases: 
(i) ingest the LSST alert stream, (ii) annotate all alerts, (iii) filter alert stream, (iv) redistribute alerts. Moreover, it goes beyond basic requirements, expanding these tasks and aiming to optimise its technology for LSST and its related Science Collaborations. Below we introduce all \fink\ requirements, marking non-traditional outputs for brokers pre-LSST era with a star ($\star$).
 
\begin{enumerate}
    \item Ingest the LSST alert stream: 
        \begin{itemize}
        \item Ingestion of alert stream batch within seconds.
        \item Big data processing robustness for the LSST alert stream volumes (inbound and outbound).
        \end{itemize}
    \item Annotate alerts: 
        \begin{itemize}
        \item[$\star$] Ability to ingest continuously updated catalogues.
        \item[$\star$] Multi-wavelength and multi-messenger event cross-matches from survey feeds.
        \item Classify alerts within minutes of observations. 
        \item[$\star$] Early light-curve classification.
         \item[$\star$] Provide classification for a subset of science cases which is continuously improved using state-of-the-art deep learning and adaptive learning techniques.
        \end{itemize}
    \item Filter:
        \begin{itemize}
            \item Allow customizable filtering of alerts. 
             \item[$\star$] Availability of added values and historical data for filtering.
        \end{itemize}
    \item Redistribute:
        \begin{itemize}
            \item Transmit annotated and reduced stream within minutes.
            \item Support forwarding of partial streams to downstream teams.
            \item Support web-interface and API clients to redistribute alerts to science teams and follow-up facilities.
              \item[$\star$] World-public access of all of our products and original alert stream (with the exception of proprietary catalogues/data).
        \end{itemize}
    \item Additional science and technology requirements:
        \begin{itemize}
             \item[$\star$] Save all raw and enriched alerts\footnote{Enriched alert packets are the original Apache Avro packets sent by LSST (or ZTF) forwarded, with new fields from the \fink\ processing. The latest schema is provided to the users via the \fink\ client. However, internally, we make use of the Parquet format which offers more flexibility in terms of post-processing.} to allow post-processing and reproducibility.
             \item[$\star$] Version control of the state of the broker and added values to reproduce selection functions.
             \item[$\star$] Possibility of ingesting and processing a simulated data stream to evaluate performance.
             \item[$\star$] Database architecture capable of handling billions of sources.
             \item[$\star$] Architecture that can be deployed in any cloud.
        \end{itemize}
\end{enumerate}

\subsection{Efficient big data processing} 
\label{section:bigdata}

In order to accommodate the paradigm change introduced by the multi-TB alert data set of LSST, \fink\ is designed to take advantage of new technological approaches based on big data tools.

\subsubsection{Cloud computing}
\fink\ handles large data volumes by distributing the data and the load over many machines. As alerts are received, the system is scaled by adding more machines which independently process the alerts using the same software. This is called horizontal scaling and it is highly successful in the big data industry.

To deploy such a processing, we use cloud computing that brings many benefits in terms of scalability of processing and cost effectiveness, fault tolerance, shareability of results, as well as increased portability and reproducibility. In \fink\ we currently operate a prototype on the VirtualData OpenStack-based cloud, a shared computing and storage infrastructure at University Paris-Saclay. For LSST processing, we will host the production service at CC-IN2P3, which also runs an OpenStack cloud and will have a local copy of LSST data that can be efficiently exploited for internal cross-match needs. CC-IN2P3 is a large scientific data centre committed to contribute to process 50\% of the raw LSST data as a satellite data processing centre during the operations phase of the LSST project.

\subsubsection{Distributed computation}

To efficiently process LSST big data volume, 
\fink\ is constructed around Apache Spark \citep{180560}, an open-source framework providing an advanced analytics engine for large-scale data processing, widely adopted in the big data industry and occasionally used in astronomy \citep[e.g.][]{2015arXiv150703325Z,Peloton2018,brahem2018astroide,2019AJ....158...37Z,2020MNRAS.493.5972H}. Further details and practical use cases for LSST can be found in \cite{2019A&C....2800305P}.

Apache Spark is based on the MapReduce cluster computing paradigm, using implicit data parallelism. It leverages the data storage distribution (data chunks) to process each chunk in parallel on different machines. When the dataset size increases, as long as it can be divided into more chunks, adding more machines result in a close-to-linear performance scaling.

One another important feature of the framework is its built-in fault tolerance. The computation is done on coarse-grained transformations that apply the same operation to many data items rather than fine-grained updates to mutable states. The fault tolerance is achieved by logging the transformations used to build a dataset (called its lineage) rather than the actual data. If a machine performing a computation is lost, the lost data can be recomputed in parallel on other machines by accessing its lineage (available from other machines).

We mainly use the Spark Structured Streaming processing engine \citep{Armbrust:2018:SSD:3183713.3190664} for building end-to-end streaming applications. Spark Structured Streaming connects seamlessly with Apache Kafka, currently used by ZTF \citep{2019PASP..131a8001P} to deliver alerts at high rates, and envisioned by the LSST alert distribution system \citep{ldm612,DMTN-093}. The combination of the two frameworks guarantees high performance streaming, with throughput tested up to hundred thousands of alerts per minutes using modest CPU and memory resources (see also Sec. \ref{section:deployment}). In \fink, incoming alerts are distributed over all the machines and they are processed independently of each other. If the pool of machines is not enough, the Spark cluster scales by adding a new machine that receives extra alerts. In the case of structured streaming, fault-tolerance is enforced through checkpointing and Write-Ahead Logs (end-to-end exactly-once fault-tolerance). If there is a partial hardware failure and data is not collected properly, the system will automatically start a new machine that will take over the stream processing where it stopped or recover missing data if alerts are still available from the distribution system.

To store incoming and processed alert data across multiple machines, we use the Hadoop Distributed File System (HDFS) capable to handle PB of data efficiently. We guarantee high availability and reliability (fault-tolerance) by replicating the data across multiple data-nodes. \fink\ can be adapted to use different storage systems and we expect to transition to the highly performing Ceph/S3\footnote{Currently, HDFS is installed on top of an existing Ceph instance, so migration would reduce the layers. Since object storage platforms are becoming widespread (e.g. Amazon S3, Ceph, Swift), they may offer the best combination of availability, long-term support and good performances in the future.}.

\subsection{Interoperability}\label{section:interoperability}

\fink\ is designed to take advantage of widely used services and tools in astronomy, thus benefiting from standards in place and ensuring communication between different surveys,  science collaborations and observational facilities. 

It has been developed to use existing reference astronomical databases for annotating information on the alert stream. For example, we have a tight integration with services developed and maintained at the Centre de Donn\'ees astronomiques de Strasbourg, including cross-matching with the latest information from survey catalogues (see also Sec. \ref{section:cross-matching} \& \ref{section:science-verif-xmatch}). We are also developing the necessary tools to cross-match alerts smoothly with other catalogues and to recover recently classified astrophysical objects (e.g. using services such as TNS).

Furthermore, the broker is capable of handling VOEvents \citep{VOE:2011} and VOEvent transport protocol \citep{VOE:2017} for reception of new multi-wavelength and multi-messenger alerts. This framework is paramount to guarantee optimal exploitation of the data gathered by LSST in real time and efficient communication with the different astronomical communities (see also Sec. \ref{section:cross-matching}). 

We seek to redistribute alerts to the community through a web-interface and API clients. Our aim is to provide reliable, customisable alert streams for science collaborations, follow-up facilities, downstream broker services and citizen science portals.

\subsection{Modularity and versioning}\label{section:modularity}

\fink's system is vastly modular, allowing fast deployment (redeploying only what has changed), fine-grained control of resources required (tuning dynamically the computational resources used by each component), and integration of new science modules with ease. This modularity also provides analysis flexibility, as we can re-run parts of the whole processing chain on-demand and only update results. All information is stored in permanent databases (incoming alerts and added-values), the software is containerised, and the analysis chains are automatised to reduce human intervention. We are transitioning from Apache Mesos to Kubernetes to improve flexibility and portability on different computing platforms, including cloud platforms such as AWS, Google Cloud and Azure, that provide this service.

\fink\ packages are structured as following: {\sc fink-broker} is the broker framework which contains the data ingestion and processing workflow (see Section~\ref{section:data-ingestion}). {\sc fink-science} contains the science modules used to annotate alerts (see Section~\ref{section:sciencemodules}). {\sc fink-filters} contains filters used to define substreams (e.g. supernovae). {\sc fink-client} is a package to collect and manipulate \fink's annotated alerts. To benchmark and test our broker, we have developed the package {\sc fink-alert-simulator} to simulate alert streams by replaying simulated or historical data.

Furthermore, we implement version control in each of our components to ease the post-processing and allow reproducibility. We use the MAJOR.MINOR.PATCH system, and both software and data are versioned. This allows traceability of our alert outputs which are necessary, for example, to reproduce selection functions. We note however that, as the number of modules in {\sc fink-science} grows, it raises questions about interface definition and migration of cross-modules (modules that need input or output from other modules). There are some standards inside Fink that have been defined and must be enforced (such as the I/O format and the field naming) when defining a new {\sc fink-science} module, but more work on the standardisation is needed to accommodate for many cross-modules over the long-term.

It is possible to update individual science modules, but the version number is global. To know what change has been made in between versions, not only we version, but we record change in a human readable way as well\footnote{\url{https://github.com/astrolabsoftware/fink-science/releases}}.

In this paper we present science verification results in Section~\ref{section:sciecenverification} using {\sc fink-broker} version 0.7.0, {\sc fink-science} version 0.3.7, {\sc fink-filters} 0.1.6, and doing simulations with {\sc fink-alert-simulator} version 0.1.1.
All components are open source and freely available\footnote{\url{https://github.com/astrolabsoftware}}, licensed under an Apache License 2. 
 
\section{Implementation}\label{section:implementation}

\fink's main structure is described in Fig \ref{figure:architecture}. Our broker is deployed in the cloud and has been tested for LSST requirements (Section~\ref{section:deployment}). The stream of alerts is continuously collected and stored on disks (Section \ref{section:data-ingestion}). A first series of filters selects good quality data that are then processed and annotated by the science modules (Section \ref{section:sciencemodules}). This annotated alert stream is stored in a database that can be used for post-processing (Section~\ref{section:db}). One key feature of \fink\ is its feedback loop to improve classification modules and to provide a recommendation system for follow-up (Section~\ref{section:feedback}). Finally, in Section~\ref{section:howto} we present how the community is able to access the broker's outputs.

For prototyping, we have used the alert stream structure envisioned for LSST, which is also adopted by the ZTF alert stream. 
All releases of the broker are benchmarked against LSST requirements and beyond in terms of alert rate. The broker has been successfully tested with up to 100,000 incoming alerts per minute using replayed ZTF alert data streams (5 times the predicted rate of alerts for LSST): generation of alert packets similar to LSST, volume stress testing, speed testing.

\subsection{Deployment platform and prototyping}\label{section:deployment}

\fink\ is a cloud-ready distributed system, currently developed and tested on the VirtualData OpenStack cloud at University Paris Saclay. The broker is made of four main clusters shown in Fig. \ref{figure:architecture}: processing (Apache Spark cluster), communication (Apache Kafka cluster), science portal (Apache HBase cluster), and data store (HDFS cluster). 

As a reference, we provide the resources deployed to prototype \fink\ in the light of LSST requirements, and currently used to process ZTF data and LSST-like simulations. The alert stream is continuously collected and processed inside a cluster of 11 machines (162 cores total for the computation, 2 GB RAM/core) with Apache Spark deployed and managed by Apache Mesos. All machines use CentOS7 and we are currently running Apache Spark 2.4.5 (N-2 released version). The associated data store is a HDFS cluster of 11 machines with 35 TB of storage total (volume cinder HDFS), and a replication factor of 3 (i.e. 100 TB of raw dedicated storage). We monitor the performances 24/7 using Grafana and Ganglia.

After processing, the stream can be divided into small substreams based on user's defined science filters. Currently, the substream data are sent to users via a cluster of 5 machines (20 cores total, 2GB RAM/core) where Apache Kafka and Zookeeper are deployed. This cluster also handles alert simulations, by replaying ZTF streams at a tunable rate. In addition, each machine has a 1 TB local disk mounted to keep these substream data available up to several days. At the end of each night, all processed data is aggregated and pushed into an Apache HBase cluster connected to the main HDFS file-system which can be accessed through the Science Portal (see Section~\ref{sec:science-portal}). We are managing different HBase tables serving different purposes: web application, light-curve post-processing and graph processing.

While we have largely enough resources to process, store and manipulate ZTF alert data, and medium-size sets of LSST simulations, we would need to expand the computational resources for LSST operations. Based on current LSST project figures, for processing LSST alert stream during the next decade we would require 500 cores total (computation, data store, and user servicing) and 1 PB disk storage per year of alerts. Thanks to the nature of cloud computing, the computing and storage resources are flexible, and we can dynamically increase (or decrease) it based on our needs.

\subsection{Data ingestion}\label{section:data-ingestion}

\fink\ ingests alert packets in real-time, decodes the alerts and automatically stores writes them to persistent storage for later re-processing needs. Alert packets can include information on the alert candidate brightness evolution over time as well as contextual information such as image cutouts and close-by sources.

Among several configuration aspects, the choice of trigger time is important (user-defined configuration option). The Apache Spark cluster pulls the trigger to Apache Kafka communication, and the time between two triggers is the result of two orthogonal goals: processing and releasing emitted data as fast as possible (small trigger time) and filling machines with as many alerts as possible per trigger to internally benefit from batch processing (large trigger time). To find the best compromise, we combine a time-based trigger with a number of alerts, and we trigger the one that comes first. Typical trigger happens after 400 alerts have been received or 15 seconds elapsed. Except the trigger, the other details of implementation are not \fink-specific. We rely on the built-in Spark-Kafka connector and the cluster manager to deal with parallelism details (how new alerts are distributed on the machines, etc.).

\begin{figure}
    \centering
	\includegraphics[width=\columnwidth]{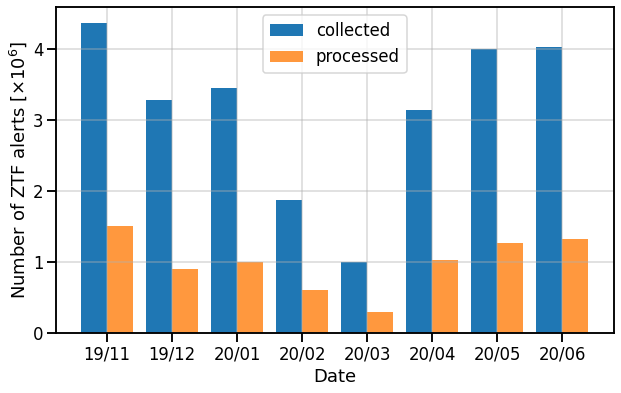}
    \caption{Monthly number of ZTF alerts collected by \fink\ (blue), and number of alerts processed by the \fink\ science modules after filtering the artefacts and bogus alerts (orange). Between 2019/11/01 and 2020/06/30, \fink\ collected more than 25 million alerts, and only about 30\% of collected alerts are satisfying the quality cuts and processed (8 million alerts).}
    \label{figure:monthlyztfstat}
\end{figure}

\fink\ is currently applied to the ZTF public stream. The ZTF alert stream is an unfiltered, 5-sigma alert stream \citep{Masci_2018}. Each night, the broker ingests in real-time the alert packets sent by the telescope. Each alert packet is serialised in Apache Avro format, and includes information on the alert candidate such as the timestamp, position, magnitude estimate, or data calibration and image cutouts around the position of the alert candidate as well as its history over a month \citep[LSST plans to extend the attached history up to a year;][]{ldm612}. The collected data between November 2019 and June 2020 represents about 130GB of compressed data per month for ZTF (about 3.1 million alerts per month on average) as shown in Fig. \ref{figure:monthlyztfstat}, and we expect between 1 and 2 orders of magnitude more alerts for LSST.

The incoming ZTF public stream is then filtered by \fink\ to reject artefacts and known bogus alerts. We use a combination of different criteria to asses the quality of alerts, partly suggested by the ZTF team:

\begin{itemize}
    \item RealBogus scores assigned by the ZTF alert distribution pipeline \citep{2019PASP..131c8002M, 2019MNRAS.489.3582D}. The values must be above 0.55.
    \item Number of prior-tagged bad pixels in a 5 x 5 pixel stamp. The value must be 0.
   \item The difference between the aperture magnitude and the PSF-fit magnitude. The absolute value must be lower than 0.1.
\end{itemize} 

This filtering reduces $\approx 70\%$ of the initial ZTF public stream, leaving $\approx30\%$ of all collected alerts to be further processed by the science modules (By only using the criterion from the ZTF real-bogus classifier, rb, 35\% of data is cut). For LSST, different criteria will be defined to select high quality alerts.

\subsection{Science modules} \label{section:sciencemodules}

Science modules in \fink\ are designed to add value to incoming alerts through cross-matching with catalogues or survey feeds (Section~\ref{section:cross-matching}) and by providing preliminary  classification scores (Section~\ref{section:classifiers}). They are external Python modules\footnote{\url{https://github.com/astrolabsoftware/fink-science}} encapsulating necessary routines and classes to process and annotate the data, and can be loaded on-demand. 

Some modules were designed specifically for \fink, others are existing user programs which were integrated into it. We aim to introduce minimal change for existing user programs to work with the broker, mainly requiring some I/O formatting. We inspect the running time of each module, optimise the entire pipeline to reduce latencies, to ensure a rapid response and an end-to-end processing time within minutes. Modules are tested individually (unit tests) and globally (integration tests) via our continuous integration pipeline that runs each time a new modification is done.
Alert throughput for currently implemented science modules are shown in Fig. \ref{figure:science_module_performances}. Some modules process all alerts that pass the quality cuts and therefore need to have high throughput (e.g. cross-match). However, some classifier modules process only alerts that satisfy additional criteria and can afford a lower throughput. 

\subsubsection{Cross-matching modules}\label{section:cross-matching}

The cross-matching modules take all the alerts that pass the quality criteria and add information on close-by sources from astronomical catalogues and other real-time surveys. Three types of cross-matching are done by \fink: (i) using the  {\sc xmatch} service provided by the Centre de Donn\'ees astronomiques de Strasbourg (CDS), (ii) using \fink\ designed cross-matching tools for small catalogues and (iii) stream cross-match using the {\sc Comet} broker.

First, \fink\ retrieves the information from external catalogues already available from the astronomical community. It currently uses the {\sc xmatch} service provided by the Centre de Donn\'ees astronomiques de Strasbourg (CDS). This service is tailored for cross-matching large catalogues, and provides an intuitive API. Alerts are sent by large batches to the {\sc xmatch} service to reduce the number of necessary calls. The service performs a positional cross-match within a radius of 1 arcsecond with the objects from the Simbad catalogue \citep{SIMBAD:2000}, and returns all results within a few seconds. 
In the case of multiple matches within this radius, we currently select only the closest object. This piece of information is then added to the original alert packet. The choice of catalogues to cross-match against, and the cross-match radius can be changed in the future but it is standardised for all alerts. In case of a failure of the external service (maintenance, or downtime of {\sc xmatch}), the broker skips this step, and continues the rest of the alert processing to ensure fast throughput for users. While subscribers will not be able to get the cross-match information in real-time, a re-processing of the missing cross-matched data on a later time can be accessed in the Science Portal.

Second, we have designed \fink\ specific tools for cross-matching with small catalogues (e.g. Section~\ref{section:SVGRB}). Third, the broker also cross-matches alerts with other ongoing alert stream surveys, such as LIGO/Virgo, Fermi or {\it Swift}. Most of these surveys are packaging alerts as VOEvents \citep{VOE:2011}, and they use the VOEvent Transport Protocol (VTP) to distribute it \citep{VOE:2017}. We use the {\sc Comet} broker 
\citep{2014A&C.....7...12S}, an open source implementation of VTP, to subscribe to these streams and receive VOEvents sent by other telescopes. Received alert packets are converted into a live stream suitable for processing, which is cross-matched with the main 
stream. The cross-match information is then added to original alerts, and distributed to the users
(see also Sec. \ref{section:filteredstreams}). We do not redistribute alerts of interest yet in the traditional VOevent format. This service is still experimental at this stage.

\subsubsection{Classifier and Anomaly Detection modules} \label{section:classifiers}

To further enrich the alert stream, these modules provide classification of the alerts using the alert stream and cross-matching information. We aim to incorporate state-of-the-art machine learning classification algorithms for a variety of science cases with three main goals.

First, we aim to obtain rapid and reliable characterisation of transient events while they are still active, enabling the selection of objects for follow-up efforts. This challenging task requires algorithms that are able to characterise objects with only a handful of light-curve observations  \citep[e.g.][]{Moller:2019,Muthukrishna:2019,Godines2019,Jamal:2020}. 

Second, we aim to enable the construction of improved training samples for these machine learning applications and provide meaningful probabilities that quantify the classification uncertainty due to shortcomings of training sets. Supervised learning methods rely on training sets known to be incomplete or not representative, biasing the resulting classification probabilities. In order to improve training samples, it is necessary to plan the acquisition of labels for objects which help increase this representativeness and introducing them to the training sets (see Section~\ref{section:feedback}). Identifying such objects is the goal of active learning algorithms which have proven to be successful in a series of astronomical applications \citep{Solorio:2005,Richards:2012,Ishida:2019}. Furthermore, Bayesian neural networks (BNNs) have been shown to provide meaningful classification uncertainties that are linked to the representativeness of training sets and can be used as information for the AL loop \citep{Moller:2019,Walmsley:2020}.

Third, we aim to provide anomaly scores whose accuracy improve with the evolution of the survey. In order to fully exploit the potential of LSST for new discoveries, \fink\ is designed to have a specific anomaly detection module, based on contemporary adaptive machine learning techniques which will be specifically designed to optimise the use of domain knowledge  \citep{Ishida_ALanomaly:2019}.

\subsubsection{Currently implemented science modules}\label{section:implementedSM}

We summarise here the currently implemented science modules in \fink. Additional ones are in development and new contributions are encouraged:
\newline

\underline{Cross-matching modules}: 
    \begin{itemize}
        \item Catalogues: Simbad catalogue \citep{SIMBAD:2000}, with a matching radius of 1 arcsecond, using the {\sc xmatch} service provided by CDS.
        \item Surveys: LIGO/Virgo, Fermi, {\it Swift} alerts via the {\sc Comet} broker (live), and survey public catalogues (post-processing).
        \item Other services: Transient Name Server (TNS) for recent classifications.
    \end{itemize}
    
\underline{Classification modules}: 
    \begin{itemize}
        \item Microlensing: Classification of events using {\sc LIA} based on \citet{Godines2019}.
        \item Supernovae partial and complete light-curve classification: 
        Recurrent Neural Network (RNN) architecture on {\sc SuperNNova} from \cite{Moller:2019}.
    \end{itemize}

Additionally we determine potential Solar System object based on a series of filters.

\begin{figure}
    \centering
	\includegraphics[width=\columnwidth]{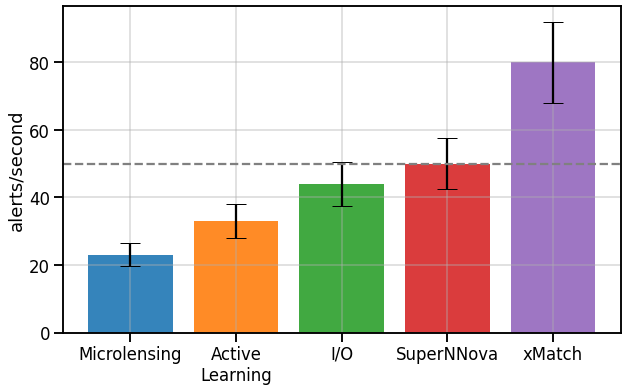}
    \caption{Current single core alert throughput (alerts/second) for each science module deployed in \fink, using replayed ZTF alert data. Microlensing: microlensing classification based on Lens Identification Algorithm \citep[LIA;][]{Godines2019}, Active Learning: supernovae classification based on Random Forest \citep{Ishida_ALanomaly:2019}, SuperNNova: supernovae classification using {\sc SuperNNova} \citep{Moller:2019}, xMatch: CDS cross-matching service. We also show the I/O throughput, that is the number of alerts per second written on HDFS by a single CPU. For reference, the grey dashed horizontal line shows a single science module throughput corresponding to 2 seconds processing using 100 CPU at LSST scale (10,000 alerts received every 37 seconds, before applying quality cuts). The total combined throughput is about 10 alerts/second/core, that is a total latency of 10 seconds to process 10,000 alerts on 100 cores. While we are working to bring all individual module performances above the threshold, the classifiers that use additional filters before processing alert data can afford a lower throughput (e.g. supernovae modules or microlensing).}
    \label{figure:science_module_performances}
\end{figure}

\subsection{Post-processing with database}\label{section:db}

The broker collects and stores all incoming alerts, as well as additional information derived by its science modules. The data is stored on the HDFS cluster (see Fig \ref{figure:architecture}), and it remains accessible for further investigations or re-processing. Given the multi-TB size of the dataset, specific tools are required to analyse it efficiently such as Apache Spark which allows real-time or post-processing analyses with little changes and on the same computing platform. Apache Spark has the advantage of being able to hold the dataset in the memory of the different machines as long as it is required for the analysis, and to combine the results globally only at the end, transparently to the user, resulting in very high performance when exploring the historical data.

All the processing tools used in live processing by the broker can be re-run on historical data collected over the years. \fink\ is thus able to quickly perform comparisons of performances of different machine learning models, adding cross-matches with other catalogues, or exploring new processing modules while keeping the development cost low. Further, all the broker tools and science modules are versioned which will be key to properly track selection functions for a variety of science cases during the LSST decade and beyond.

\subsection{Feedback}\label{section:feedback}

One of the defining features of \fink\ is its designed ability to improve the accuracy of added values as the survey evolves by enabling the use of adaptive learning strategies. Each science module which makes use of such strategies encapsulates the learning loop within it. 

To improve the accuracy of specific added values, potentially informative objects are identified in the stream and their labels are searched continuously in known public data bases using our cross-matching module. Moreover, \fink\ will make public the list of objects of interest for each science module at least via two channels: the Fink client (live Kafka streams), and the Science Portal (web application). We aim to ensure that information is spread through the spectroscopic follow-up community through standard channels such as TNS for transient discovery, VOEvents for multi-messenger counterparts and Target of Observation Manager (TOM\footnote{\href{https://lco.global/tomtoolkit/}{https://lco.global/tomtoolkit/}}) systems interfacing between brokers and follow-up facilities.

Once a new informative label is available, the user is notified and the machine learning models can be retrained. This process can be computationally very expensive depending on the volume of training data, complexity of the model and frequency with which labels are provided. For the first two science modules employing adaptive learning strategies (SN and Anomaly Detection) \fink\ resources will be available for automatic model update. Other community projects willing to implement such strategies can take advantage of the infrastructure provided by \fink\ for cross-match and advertisement of desired labels but will be initially responsible for retraining their own modules. Further arrangements to automatise this process within the broker will require an evaluation of the computational cost and available resources.

The constant update of the machine learning models also means that \fink\ will be able to reprocess previous alerts and provide more accurate classifications of historical with the evolution of the survey. We plan to hold frequent data releases with updated classifications and anomaly scores which can highly improve the applicability of catalogue data in photometric studies.

\subsection{How to use \fink\ }\label{section:howto}

The annotated alerts by \fink\ can be accessed by the community through live filtered streams (Section~\ref{section:filteredstreams}) and a web portal (Section~\ref{sec:science-portal}).

\subsubsection{Live filtered streams}\label{section:filteredstreams}

Users can receive a customised filtered alert stream at the end of the processing, when all the additional information from the science modules has been computed. We aim to provide these filtered streams within minutes of the alerts being ingested.

A filter uses the information available in the alerts and the added values provided by the broker
(e.g. cross-matches, classification scores) to select a subset of the alert stream for a given science case\footnote{These filters can be simple boolean expressions, or more complex workflows, but they do not add additional information.}. The only criterion for a filter to be deployed
is that the volume of data to be transferred is tractable on our side. Hence each filter must focus on a specific aspect of the stream to reduce the outgoing volume of alerts.

The alerts are distributed using Apache Kafka (see Fig. \ref{figure:architecture}), and Kafka topics are created based on the user-defined filters. Users are encouraged to register to connect to existing streams, or create new ones\footnote{\url{https://fink-broker.readthedocs.io/en/latest/fink-client/}}. As we are in testing operations phase, currently filters are checked and deployed by our local
team. We keep the sub-stream data available for 7 days 
after emission, after which it is deleted (but all processed data by \fink\ remains available in the Science Portal). We are constructing an API service for interested teams, enabling follow-up facilities and downstream brokers to create and receive automatically a customisable reduced alert stream.

\begin{figure}
\centering
\includegraphics[width=\columnwidth]{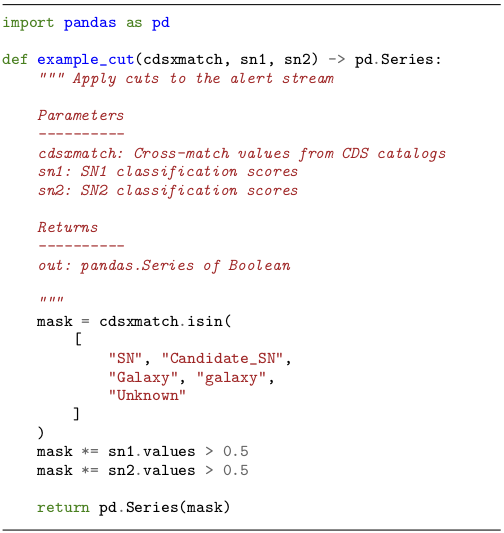}
\caption{Example of a very simple filter in Python to select possible supernovae using CDS cross-match and the two supernova classifiers currently implemented in \fink. An optimised filter is further discussed in the science verification section.}
\label{figure:filter_example}
\end{figure}

An example of a filter is given in Figure \ref{figure:filter_example}. For this particular filter, the original stream for a ZTF observing month of $1,515,953$ alerts (after quality cuts) was reduced to $112,344$ alerts. This simplistic filter is enhanced and discussed in Section \ref{section:sciecenverification}.

\subsubsection{Science portal}\label{sec:science-portal}

The \fink\ Science Portal allows access and manipulation of the processed data through a web interface. The data is grouped by object ID\footnote{There is a unique identifier for all alerts corresponding to the same astrophysical object based on the sky position. For LSST, we will use what the project provides \citep{ldm612}.}, providing a full alert history of an object at the end of each night beyond the limited history provided by the alert stream.

\begin{figure*}
    \centering
	\includegraphics[width=\textwidth]{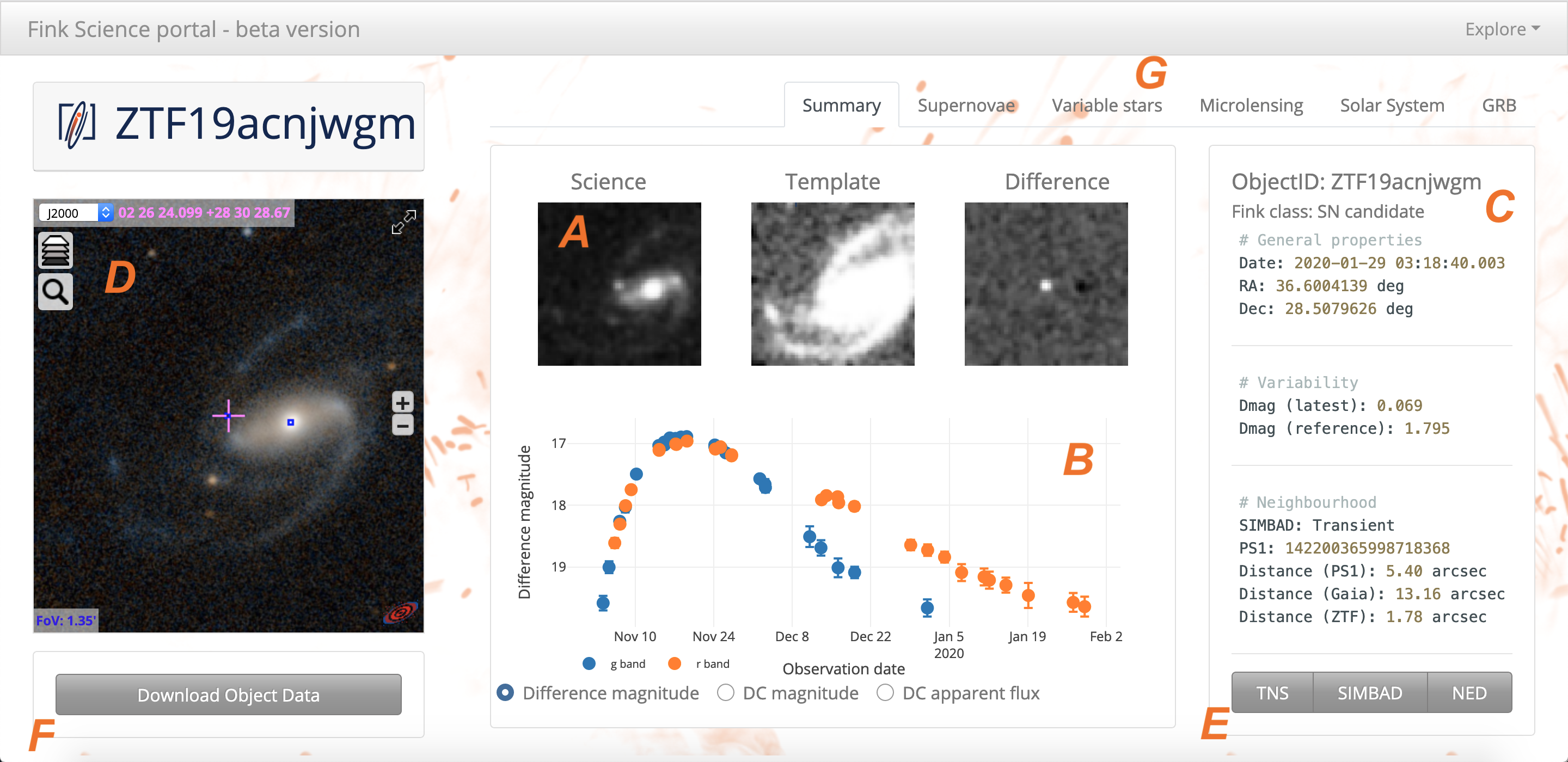}
    \caption{The \fink\ object page summary from the beta version of the science portal (\url{https://fink-broker.org}). \textbf{A}: Cutouts from the last alert packet of this object. \textbf{B}: Complete light-curve of the object. The users can change units to display difference magnitude, DC magnitude, or DC apparent flux as a function of time. Filter bands are shown in different colours. \textbf{C}: Textual information from the last alert, such as its emission date, sky position, light variability with respect to the previous alert, or the closest sources from external catalogues (PanSTARRS-DR1, Gaia-DR2, ZTF). \textbf{D}: Interactive Aladin view of the object with the PanSTARRS-DR1 image centred at the position of the latest alert. \textbf{E}: External online data-sets (TNS, SIMBAD, NED) with cross-matched data to this object (e.g. spectroscopy). \textbf{F}: Button to download object data (all annotated data for this object). \textbf{G}: Tabs to trigger different views of the object. The views focus on specific aspect of the data (such as the evolution of classification module scores for Supernovae detection) and the users can perform data fit based on pre-loaded models (e.g. variable stars, microlensing). New features and new views are regularly added based on our community feedback. As of November 2020, the Science Portal gives access to more than 15 million ZTF alerts processed by \fink.}
    \label{figure:science-portal}
\end{figure*}

During the processing, alert data is primarily stored in the Apache Parquet format which seamlessly integrates with our processing tools. But to make the data exploration of end-user products as quick and relevant as possible given the expected PB of alert data at the end of the survey, we use Apache HBase for the Science Portal data storage (key-value storing). Apache HBase is a non-relational distributed database that allows to easily keep track of changes, to perform fast queries, and it allows flexibility in the data model schema. An HBase cluster has been deployed in the VirtualData cloud, data is stored on HDFS (see Fig. \ref{figure:architecture}). The system
currently uses an HBase-Spark connector to ingest and update data at the end of the night, and the front-end to access the distributed data, deploy jobs to data, move data between servers is under development. A beta version of the Science Portal (i.e. with limited features) is online at \url{https://fink-broker.org}, and Fig. \ref{figure:science-portal} describes the main object view page summary. 

Note that the user data intensive tasks will be performed in the \fink\ Science Platform, which will use Jupyter notebooks, and has \fink\ specific-tooling abstracting Apache Spark to allow scientists to manipulate large volumes of data without the need of learning Spark. The Science Platform is not yet publicly available as we are still sizing it and trying to find the best compromise between flexibility and security.

Alongside the main key-value storing scheme, we are developing\footnote{This project is part of the \href{https://hepsoftwarefoundation.org/gsoc/2020/proposal_AstroLabBigDataGraphs.html}{Google Summer of Code 2020}} a graph database to explore differently and efficiently alert data collected by the broker.

\section{Science Verification}\label{section:sciecenverification}

\begin{figure*}
    \centering
	\includegraphics[width=\textwidth]{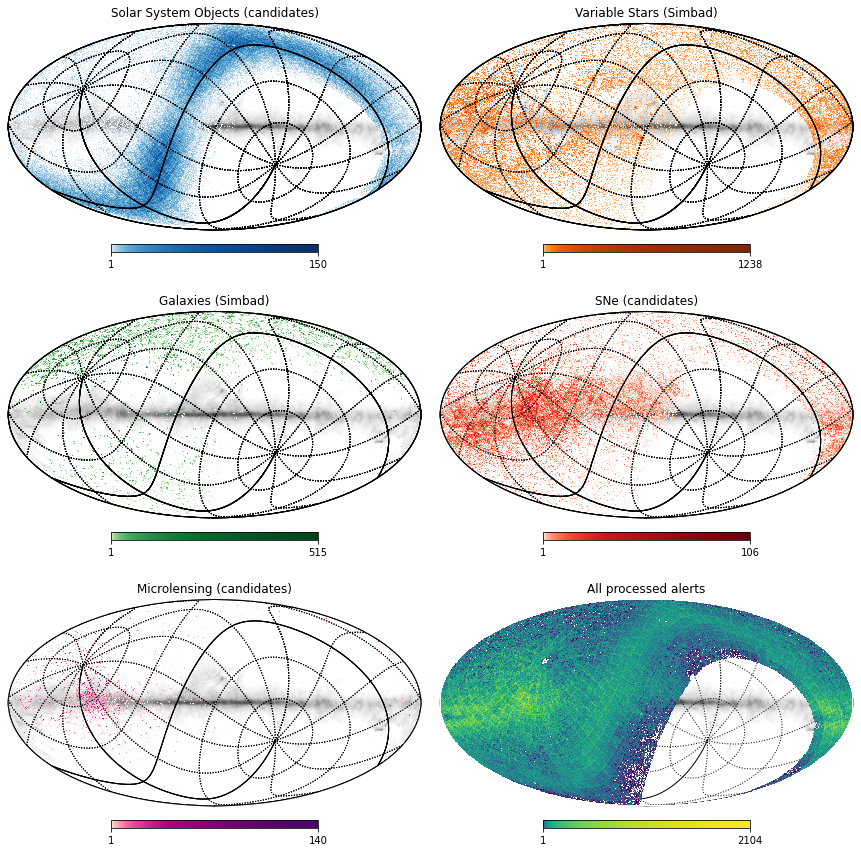}
    \caption{Footprint of the ZTF alert stream from November 2019 to June 2020 associated to a subset of transient types using current \fink\ science modules: candidate Solar System objects (top-left blue, see Sec. \ref{section:sso}), variable stars from the cross-match with the Simbad catalogue (orange top-right, see Sec.~\ref{section:science-verif-xmatch}), alerts matched to a galaxy in the Simbad catalogue (green middle-left, see Sec.~\ref{section:cross-matching}), supernovae type Ia candidates selected using SuperNNova (red middle-right, see Sec.~\ref{section:SN2}), microlensing event candidates selected using LIA (purple bottom-left, see Sec.~\ref{section:science-verif-microlensing}), and all 7,975,588 processed alerts by \fink\ that pass quality cuts (bottom-right, see Sec.~\ref{section:data-ingestion}). The Planck Commander thermal dust map \citep{Akrami:2018mcd} is shown in the background for reference. All maps are in the Galactic coordinate system, with a healpix resolution parameter equal to \texttt{Nside=128} \citep{Gorski_2005}, except for alerts matching galaxies (green middle-left) where \texttt{Nside=64} has been used to increase the readability.}
    \label{figure:footprint}
\end{figure*}

We are developing \fink\ using replayed ZTF alert data since November 2019. In February 2020, the \fink\ collaboration signed a memorandum of understanding (MoU) with ZTF to receive the live public data stream and has been since then performing technical and science tests with its data. In this Section we showcase our first science modules and alert selection methods for a handful of science cases. 
Results have been bench-marked using, when relevant, simulations and the ZTF public stream (alert data from November 2019 to June 2020). 
We show the footprint of the ZTF alert stream in Figure~\ref{figure:footprint} and alerts associated to a subset of transient types using our current science modules.

\subsection{SIMBAD database cross-matching module}\label{section:science-verif-xmatch}

Each alert that passes our broker's quality cut criteria (see Section~\ref{section:data-ingestion}) is cross-matched against the Simbad catalogue using the {\sc xmatch} service provided by CDS (see Sec. \ref{section:cross-matching}). On each processor, the alerts are sent by batches to minimise the call to the service, and we currently achieve a throughput of about 80 alerts/second on a single processor (including network latencies, and cross-matching time), which corresponds to about 1.25 second to process 10,000 alerts with 100 processors (expected rate of alerts for LSST every 37 seconds, see Fig. \ref{figure:science_module_performances}). 

\begin{figure}
    \centering
	\includegraphics[width=\columnwidth]{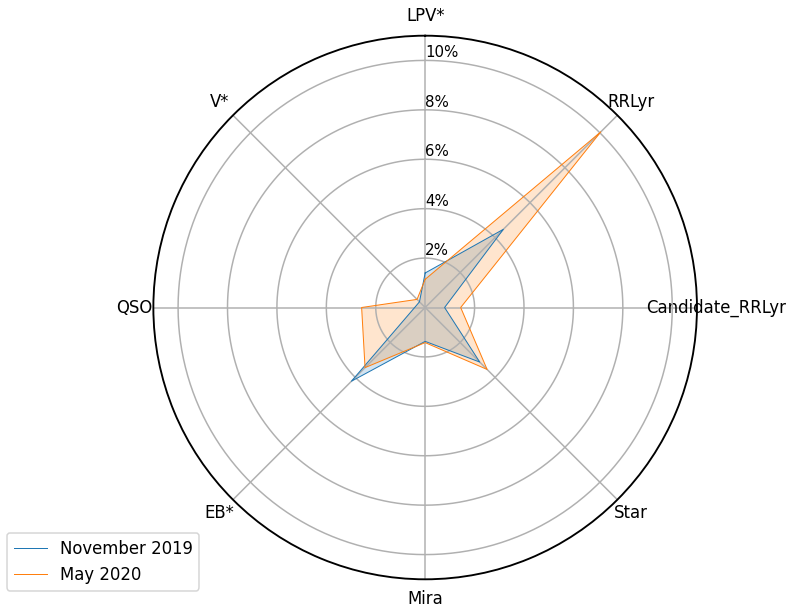}
            \caption{ZTF alert stream classification returned by the {\sc xmatch} service at CDS, using the Simbad catalogue. We show the results from the 8 main categories (LPV*=Long-period variable star, RRLyr=Variable Star of RR Lyr type, Candidate$\_$RRLyr=Possible Star of RR Lyr type, Star=confirmed star, Mira=Variable Star of Mira Cet type, EB*=Eclipsing binary, QSO=quasar, V*=Star suspected of Variability) returned for November 2019 in blue (1,515,953 alerts processed after quality cuts) and May 2020 in orange (1,276,311 alerts processed after quality cuts). Percentages are computed with respect to the total alerts processed each month.}
    \label{figure:simbad-cross-match}
\end{figure}

Using this module on the ZTF public alert stream, we find that only  $\approx$20-25\% of processed alerts are known transients or have an astrophysical object match in Simbad. The remaining are either new transients, transients not previously in this catalogue or spurious detections. The most common cross-matched types are RR Lyrae variable stars (RRLyr) and eclipsing binary stars (EB*), representing up to 10\% of the average alert stream as shown in Fig. \ref{figure:simbad-cross-match}. Each month, we have around 150 different object types returned by the cross-match service, ranging from extra-solar planets to variable stars or AGN. 

\subsection{GRB search using Fermi and Swift} \label{section:SVGRB}

The GRB module implemented within \fink\ is devoted to promptly identify the optical counterparts of GRBs in the ZTF alert stream. To do so, two methods can be independently conducted, either by performing targeted searches for ZTF candidates as soon as a GRB is detected and localised by gamma/x-ray satellites or by making blind searches for optical afterglow-like emissions in the incoming ZTF alert stream to catch untriggered (no gamma/x-ray detection) or orphan GRBs. In this Section we show results only for the targeted search since the blind search is currently being developed.

Our targeted search for GRB optical afterglows consists in quickly isolating credible transient candidates in the ZTF alert stream that are coincident in space and time with known GRBs. We select the ZTF alerts that first satisfy our standard quality cuts (see Section~\ref{section:data-ingestion}), plus we apply additional cuts (cuts0):
\begin{itemize}
    
    \item The ZTF candidates are uncatalogued sources according to the Simbad Astronomical Database. We require that ZTF candidates are at least 1 arcsecond away from any Simbad objects.
    
    \item We require that the ZTF candidates are at least 5 arcsecond away from any of the objects listed to the Gaia-DR2 \citep{Gaia16,Gaia18} and PanSTARRS-DR1 \citep{Chambers16} catalogues. 
    
\end{itemize}

We applied our selection criteria to find the remaining ZTF candidates that might be associated in position and time with a given GRB (cuts1):
\begin{itemize}
\item The ZTF alert detection time $T_{ZTF}$ must be within a searching time window $\Delta T_s = T_{GRB}-T_{ZTF} \in [-600~s;20~days]$, where $T_{GRB}$ is the GRB trigger time. This time window is designed to catch all the possible optical counterparts that may occur from GRB precursor emissions up to the supernova components that may emerge several days after the $T_{GRB}$ in the case of long GRBs.
 \item The ZTF candidates must be newly discovered transients by the ZTF survey (detections prior to $\Delta T_s$ are therefore rejected) to avoid contamination by variable sources or long lasting transient sources. 
 \item The angular separation, $D_A$, between the ZTF alert and the GRB positions must be consistent with the localisation region of the GRB, $\sigma_{GRB}$. This $\sigma_{GRB}$ is highly variable from a GRB to an other one depending on the instrument that performed the localisation.
\end{itemize}
We tested the \fink\ GRB module on {\it Swift} and Fermi-GBM GRBs detected from November 01, 2019 to June 17, 2020. During this period, a total of 46 GRBs were detected by the {\it Swift}-BAT gamma-ray instrument and well-localised to a $\sigma_{BAT}\sim2$ arc min accuracy (up to $\sigma_{XRT}\sim2$ arcsecond accuracy with the {\it Swift} x-ray XRT telescope) while a total of 142 Fermi-GBM GRBs were detected and localised with a median error radius $\sigma_{GBM}=4.13^{\circ}$. We first applied the quality cuts0 on the whole ZTF alert stream as a preliminary filter. Before applying our selection criteria (cuts1), we set up the maximum angular separation between the positions of the ZTF events and our selected GRBs to $D_{A,{\it Swift}} = 3\times \sigma_{BAT,XRT}$ and $D_{A,Fermi} = min(\sigma_{GBM},5^{\circ})$, for {\it Swift} and Fermi GRBs respectively. The results of our different selection cuts (cuts0 and cuts1) on the monthly ZTF alert stream are summarised in Table \ref{tab:GRBcuts}.

\begin{table}
	\centering
	\caption{Results of the application of our quality cuts (cuts0) and selection cuts (cuts1) on the ZTF public alert stream from November 01, 2019 to June 17, 2020 in order to find credible ZTF optical afterglow candidates associated in space and time with the Fermi-GBM and {\it Swift} GRBs. We did not find any credible candidates for {\it Swift} during this period after all cuts.}
	\label{tab:example_table}
	\begin{tabular}{lccr} 
		\hline
		Period & Raw ZTF & Raw+cuts0 & Raw+cuts0+cuts1\\
		           &    alerts ($\times 10^{6}$) & ($\times 10^{3}$) & \\
		\hline
		\hline
		Nov 2019 & $\sim4.4$ & $\sim13.0$ & 335/0 (Fermi/{\it Swift})\\
		\hline
		Dec 2019 & $\sim3.3$ & $\sim2.4$ & 23/0 (Fermi/{\it Swift})\\
		\hline
		Jan 2020 & $\sim3.4$ & $\sim4.8$ & 27/0 (Fermi/{\it Swift})\\
		\hline
		Feb 2020 & $\sim1.9$ & $\sim3.3$ & 0/0 (Fermi/{\it Swift})\\
		\hline
		Mar 2020 & $\sim1.0$ & $\sim1.2$ & 0/0 (Fermi/{\it Swift})\\
		\hline
		Apr 2020 & $\sim3.1$ & $\sim3.9$ & 0/0 (Fermi/{\it Swift})\\
		\hline
		May 2020 & $\sim4.0$ & $\sim7.5$ & 7/0 (Fermi/{\it Swift})\\
		\hline
		Jun 2020 & $\sim1.7$ & $\sim1.7$ & 1/0 (Fermi/{\it Swift})\\
		\hline
	\end{tabular}
	\label{tab:GRBcuts}
\end{table}

We found no robust association between the ZTF transient alerts and the {\it Swift} GRBs. On the contrary, because the Fermi-GBM localisation accuracy is much worse than the one of the {\it Swift} instruments, we found associations for 393 ZTF transient candidates with 22 Fermi GRBs in total over the entire period of our investigations. We visually inspected these candidates, comparing real afterglow light-curves with ZTF observations (detection+upper limits prior to the detection) to further select candidates. After applying this selection procedure to ZTF observations, of those remaining candidates with known GRB afterglow light curves (seen on-axis) we were able to rule out 97\% of them as being credible optical afterglow candidates. The remaining 3\% of the candidates (13/393 candidates) were detected by ZTF few hours up to 1.2 days after the GRB trigger times at a median r-band magnitude r = 19.61. According to the ZTF  upper limits on their optical flux obtained days before the detection and due to the ZTF survey strategy (revisit at daily timescale) which prevents from obtaining late photometric measurements at deeper magnitude ($r_{lim}\sim20.5$), we could not completely rule out their potential association with the Fermi-GBM GRBs.\\

Our preliminary implementation already allows us to make fast off-line cross-matches between the ZTF and {\it Swift}/Fermi GRB alert streams. Future implementations will target real-time identification of ZTF optical counterparts from any GRB detected by the {\it Swift} and Fermi missions, and later the SVOM mission. Additional functionalities will also allow to rapidly estimate the likelihood of the ZTF transient candidates with typical GRB afterglow behaviours depending on the viewing angle with respect to our line-of-sight. This likelihood estimator will be used to classify the ZTF candidates in different GRB sub-streams for which we could apply different follow-up and analysis strategies.

\subsection{Microlensing events}\label{section:science-verif-microlensing}

The microlensing classification module is based on the Lens Identification Algorithm (LIA) presented in \citet{Godines2019}. In short, a Random Forest algorithm is trained with simulated light-curves similar in cadence and noise to the associated survey of interest (currently ZTF). More than 50 features are extracted from the photometric light-curve and re-projected with a Principal Component Analysis to classify four different classes of events, which include microlensing. Since the current algorithm does not support multi-bands information, each individual filter-band is treated separately with its own training set and noise model. To derive the final class probability, results from each band for the same object are compared. A multi-band version of LIA is under development.

For this implementation, we simulated 10,000 microlensing events over real ZTF light-curves from the publicly released ZTF DR2 database. For each light-curve, the impact parameter, $u_0$, has been randomly chosen between 0 and 2, the maximum magnification time $t_0$ between the time of the first and last observations, and the Einstein event duration $t_E$ between 1 and 500 days, following a logarithmic distribution. We split this data set into a training set (7,500 light-curves) and a validation set (2,500 light-curves). We train LIA with the training set, and apply the model on the validation set to evaluate performance. We find a classification accuracy of ~80\% for light-curves that have at least 5 points during the magnification peak (within the $[t_0-t_E, t_0+t_E]$ time interval). In addition, ~60\% of the correctly classified events were discovered before the end of the transient (defined as $t_0+t_E$ in this test).

Additionally, to mimic ZTF alert content which provides historical data for only a month, we reduce the ZTF DR2 database light-curves to 30 days. We retrain the model using these partial curves, and apply the model on the data. We find that only 20\% alerts are flagged as microlensing before $t_0+t_E$. We note also that 50\% of correctly classified alerts are identified before the peak of the event. The drop in accuracy compared to the full light-curve case described above is expected as we have less information for each event, making hard to determine the baseline flux and its dispersion. For LSST, as each alert is expected to contain up to a year history, we expect a higher accuracy due to a better characterisation of the pre-event and event data.

We also tested this module using the ZTF public alert stream, since some microlensing events have already been found during the first years \citep{Mroz2020}. We analysed the alert data taken between November 2019 and June 2020. Among the 10 million alerts that pass the quality cuts, we classified alerts with more than 10 valid measurements per band, resulting in 41,913 alerts classified as microlensing candidates by LIA in both g and r filter bands (candidate rate around 0.4$\%$). The distribution of candidates on the sky is mainly around the galactic plane, where it is most expected due to the higher density of visible objects, as shown in Figure~\ref{figure:footprint}. We further select alerts by applying additional selection criteria:

\begin{itemize}
\item the alert must not be classified as SN Ia candidate by \fink,
\item the total history of the object must be less than 50 measurements (to select alerts corresponding to recent events only),
\item the Star/Galaxy score of closest source from the PanSTARRS-DR1 catalogue must be above 0.7 (star),
\item we keep candidates on low Galactic latitudes ($|b|$ < 20 degrees).
\end{itemize}

\begin{figure}
    \centering
    \includegraphics[width=\columnwidth]{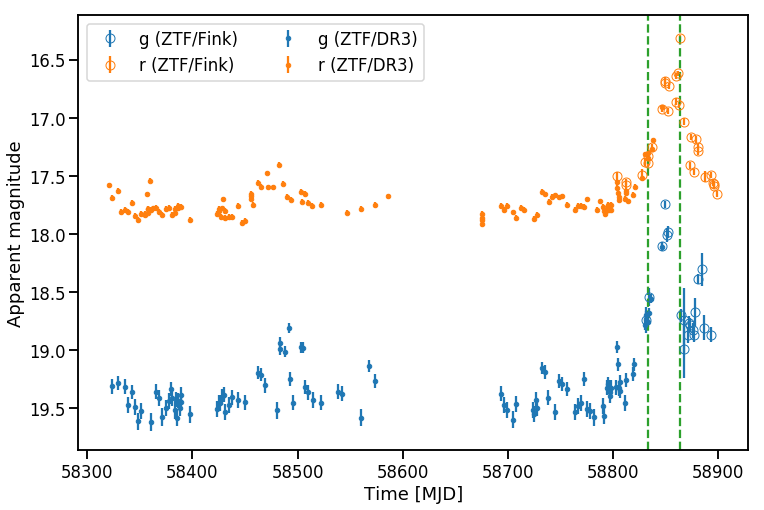}
            \caption{Lightcurve of a microlensing event candidate classified by \fink\ (ZTF object ID: ZTF18acvqrrf). Blue markers and orange markers indicate the g and r bands, respectively. Large circle markers and small circle markers indicate the Fink and ZTF DR3 data, respectively. There is a small overlap between the two datasets, but DR3 data stops at MJD=58846.2608449, where \fink\ data take over alone. The two vertical dashed green lines enclose alert data that lead to the first classification as microlensing by \fink. Error bars are estimated by the photometric algorithm only, probably leading to underestimation of errors here and partly explaining the high $\chi^2$ values in Table~\ref{tab:microlensing}.}
    \label{figure:microlensing_lightcurve}
\end{figure}

We are left with 6 alerts representing 4 unique objects (timestamp of first classification as microlensing): ZTF18acvqrrf (2020-01-12), ZTF20aaaacan (2020-01-26), ZTF20aauqwzc (2020-05-11), ZTF20aazdsjr (2020-06-22). Among the 4 objects, ZTF20aazdsjr appears to be a supernova type Ia not correctly classified by \fink\ at the moment of the alert, and ZTF20aaaacan does not appear as a microlensing event. 
We show the light-curve for ZTF18acvqrrf in Figure~\ref{figure:microlensing_lightcurve}, and the results for a microlensing point-source point-lens rectilinear model fits in Table~\ref{tab:microlensing}, including or not the data from the recent ZTF DR3 release.
Unfortunately the constraints are not significant, given the data collected, and we cannot conclude on the microlensing nature of this event as well as of the other remaining event (ZTF20aauqwzc).

\begin{table}
    \centering
    \begin{tabular}{lcrrr}
        Event & $t_0$ (MJD) & $t_E$ (day) & $u_0$ & $\chi^2$/dof \\
        \hline
            ZTF18acvqrrf & 58854.5 $\pm$ 0.1 & 14 $\pm$ 30 & 1 $\pm$ 4 & 22.4 \\
            +DR3 & 58854.8 $\pm$ 0.6 & 23 $\pm$ 7 & 0.6 $\pm$ 0.3 & 14.8 \\
    \end{tabular}
    \caption{Fit results for the microlensing event candidate ZTF18acvqrrf} discovered by \fink, using pyLIMA \citep{2017AJ....154..203B} on the simple PSPL model (1-sigma error quoted). The fit combines data for the 2 filter bands g and r. We also perform a fit using the available ZTF DR3 data (second line). In all cases, the constraints are not significant, and we cannot conclude on the microlensing nature of this event.
    \label{tab:microlensing}
\end{table}

A candidate rate of 0.1\% is still too high for LSST (this would correspond to 10,000 candidates per night before selection cuts), and we are working on a better selection of candidates as well as better classification models. In particular, the complete knowledge of the pre-event light-curves will allow to limit the search for microlensing within the sub-sample of stars that have been completely stable since the beginning of the survey, and therefore drastically reduce the false positive rate. The multi-band version of the LIA should also significantly improve the classification quality. We also plan to periodically re-classify objects using aggregated object data over time.

\subsection{Supernovae}

\fink\ obtains classification scores for potential supernovae at any stage of its evolution using the Deep Learning framework {\sc SuperNNova} \citep{Moller:2019} introduced in Section~\ref{section:SN2}. We train this classifier on simulations from \citep{Muthukrishna:2019} and test it with an independent subset of this simulation. The performance of our classification module is further evaluated using data from the ZTF public stream. For reproducibility, we make available the alerts and code used for results in this section \footnote{\fink\ annotated alerts and TNS SNe are available at \href{https://doi.org/10.5281/zenodo.4036589}{ DOI:10.5281} and to reproduce results use this \href{https://github.com/astrolabsoftware/fink-usecases}{jupyter notebook}.}.

We are also currently developing another supernova module that takes advantage of an Active Learning approach for early classification. We introduce this method 
in Section~\ref{section:SN1}.

\subsubsection{Supernova science module} \label{section:SN2}
The supernova module uses the classification scores of {\sc SuperNNova} \citep{Moller:2019}. {\sc SuperNNova} is a Deep Learning framework designed to classify light-curves using photometric data only (fluxes and errors in different band-passes) thus does not require feature extraction or pre-processing. This classifier is able to provide classification scores at any light-curve stage, including early SNe.

\begin{figure}
    \centering
	\includegraphics[width=\columnwidth]{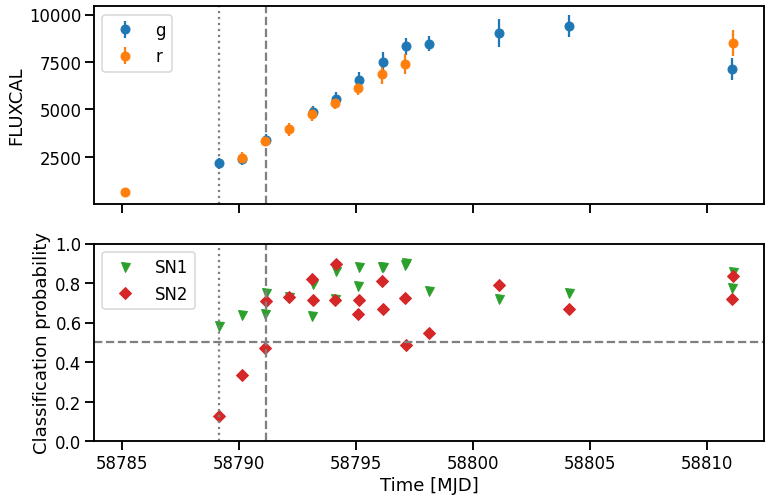}
            \caption{\textit{Top:} Light-curve of a supernova type Ia event classified by \fink\ (ZTF object ID: ZTF19acmdpyr, IAU Designation: SN 2019ugu). Blue circle markers and orange circle markers indicate the g and r bands, respectively. Error bars come from the photometry. This supernova was first reported in TNS by the broker AMPEL \citep{ampel} on 2019/11/06 (MJD 58793). \textit{Bottom:} Evolution of the classification probabilities as a function of time using {\sc SuperNNova} \citep{Moller:2019}: the first model (SN1, green triangle markers) disentangles type Ia SNe vs. non-Ia SNe, and the second model (SN2, red diamond markers) classifies a general supernova class (SNe Ia and Core-Collapse) vs. non-SNe events. \fink\ classification using only SN1 probability value above 0.5 would have happened on 2019/11/02 (dotted vertical line, MJD 58789), while using both SN1 and SN2 probability values above 0.5 would lead to a classification on 2019/11/04 (dashed vertical line, MJD 58791). This supernova was classified particularly early by \fink, and more information about the classification delays are shown in Fig~\ref{figure:snia_delay}.}
    \label{figure:snia_lightcurve}
\end{figure}

\begin{figure}
    \centering
    \includegraphics[width=\columnwidth]{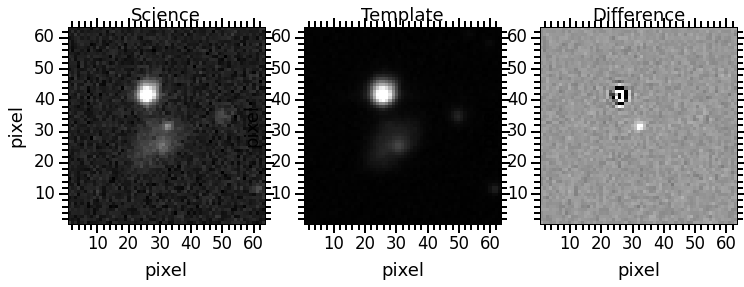}
            \caption{Alert cutouts corresponding to the supernova type Ia event shown in Fig~\ref{figure:snia_lightcurve} at the time of first classification by \fink: the observation (left), the reference image used in the subtraction (middle), and the difference image (right). The bright object on the top-left of the supernova is a foreground star. We use cutouts to visually inspect candidates.}
    \label{figure:snia_cutout}
\end{figure}

We train two {\sc SuperNNova} models using ZTF-realistic light-curve simulations. The first model, SN1, was trained to disentangle type Ia SNe vs. non-Ia SNe. When evaluated using simulations, SN1 has a classification accuracy of $75.17$ for complete light-curves and $58.80$ for classification before maximum light of the light-curve. The second model, SN2, classifies a general supernova class (SNe Ia and Core-Collapse) vs. non-SNe events and is able to obtain accuracies up to $87.07$ for complete light-curve classification.

We apply these classifiers to $2,417,284$ ZTF alerts from November and December 2019 that pass quality cuts. Selection is done filtering on added values from the alerts, supernovae classification modules and the cross-matching module.

Performance is evaluated on the ZTF alert stream by computing the reduction of the total alert stream and the number of spectroscopically supernovae selected. We identify spectroscopically classified supernovae by cross-matching the coordinates of reported TNS SNe with the alert stream requiring a match within $1.5$ arcseconds. We highlight that spectroscopically classified supernovae are not a complete sample and thus only partially indicative of our performance.\\

We select our samples by performing the following cuts: 
\begin{itemize}
    \item Deep Learning Real Bogus score $drb>0.5$.
    \item Cross-match with SIMBAD: the alert must be associated with a galaxy, a candidate transient or tagged as unknown.
    \item At least $3$ detections in the alert package are required to promote reliable classification scores ($nalerthist$ added value from 
    \fink).
    \item The alert must obtain a classification score $>0.5$ (either from SN1 or SN2).
    \item To filter long-term variable objects we require less than 400 detections in the ZTF survey ($ndethist$).
    \item To filter variable stars not present in our catalogues, we require a Star/Galaxy score by SExtractor $>0.4$.
\end{itemize}

We apply these requirements to the ZTF alerts stream and obtain a selected sample as shown in Table~\ref{tab:SNZTF}. An example of a SNIa classified by fink is shown in Figures~\ref{figure:snia_lightcurve},\ref{figure:snia_cutout}. We are able to reduce the alert stream up to $8\%$ while maintaining $>75\%$ of classified SNe Ia and SNe in the sample. Those SNe which are not selected are visually found to have only a handful of photometric epochs or to be at the end of their visible variability (tail of the light-curve). 

\begin{figure}
    \raggedleft
    \includegraphics[width=1.02\columnwidth]{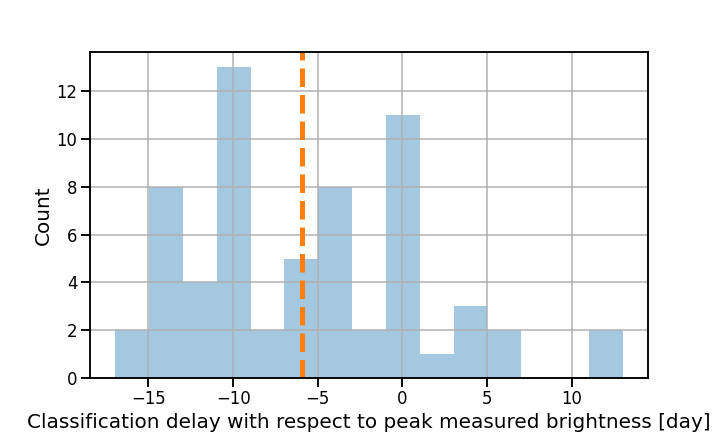}
            \caption{Delay (in days) between the first classification by \fink\ of a supernova type Ia event (using the SN1 model) and its peak measured brightness, between 01 November 2019 and 31 December 2019. Negative delays mean the classification happens before the supernova reaches its peak brightness. We show only data for supernovae type Ia that exploded after 01 November 2019 (from TNS query), and for which a peak can be identified. The median delay is -6 days (orange dotted vertical line).}
    \label{figure:snia_delay}
\end{figure}

For spectroscopically identified SNe Ia in the alert stream we estimate the delay of classification with respect of the observed peak brightness (maximum flux measured by ZTF). We obtain a median delay of $6$ days before observed peak brightness for SN1. As shown on Figure~\ref{figure:snia_delay}, SN1 is able to identify SNe Ia well before observed peak for this sample. Such an ability will be particularly relevant for coordinating follow-up efforts in the era of LSST.

These selected samples can be further reduced by applying additional cuts or increasing the classification threshold as can be seen in Table~\ref{tab:SNZTF}. Through visual inspection, we find that a large number of these alerts present some type of variability consistent with other transient phenomena or have a reduced number of photometric epochs ($<10$). Subsequent work will strive on improving science modules and filtering techniques to raise the purity and efficiency of our selection.

Future improvements to this science module include expanding pre-trained models to other SN and transient types, including new core-collapse templates e.g. \citep{Vincenzi:2019}, improving the ZTF survey simulation and implementing Bayesian NNs available in {\sc SuperNNova} to obtain classification scores with meaningful model uncertainties.


\begin{table}
    \centering
    \begin{tabular}{ccccc}
        sample & \# alerts & \% alerts & \# unique & \# unique \\
         &  & & SNe Ia & SNe \\
        \hline
        quality cuts&  2,417,284 & 100\% & 296 & 366 \\
        selection cuts&  576,190 & 23.84\% & 258 & 319 \\
        SN1>0.5 & 365,228 & 15,11\% & 242  & 296\\
        SN2>0.5 & 208,978 & 8.65\% &  223 & 275 \\
        SN1>0.6 & 308,822  & 12.78\% & 229  & 278 \\
        SN2>0.6 & 145,736 & 6.03 \% &  196 & 245 \\
    \end{tabular}
    \caption{The effect of cuts on the ZTF alert stream between November and December 2019. Columns indicate cut type, number of alerts selected, percentage of alerts for reference, number of unique SNe Ia and SNe recovered (from TNS query). First, alerts that satisfy our standard quality cuts (see Section~\ref{section:data-ingestion}). Then selection is done by applying SN-filtering specific cuts (selection cuts) defined in Section~\ref{section:SN2} combined with a threshold on either the classifier model SN1 or SN2 (e.g. $SN1>0.5$).}
    \label{tab:SNZTF}
\end{table}

\subsubsection{Supernova science module with Active Learning} \label{section:SN1}

We are currently developing a supernova classification module which uses a Random Forest (RF) classifier \citep{breiman2001} coupled with an uncertainty sampling active learning strategy to construct an optimised training sample. The module follows the paradigm established by \citet{Ishida:2019}, but employs a feature extraction method based on on a sigmoid function to model raising light curves (Leoni \textit{et al.}, \textit{in prep}). 
 
We tested a preliminary implementation on ZTF simulations \citep{Muthukrishna:2019} consisting of 
$12,560$  objects of which $7,773$
(i.e. $ \approx 62 \% $ of the total)
are SN Ia. The learning loop started from an initial training sample of 10 objects (5 of which were SNe-Ia). After 2000 iterations where batches of 1 object each were added to the training sample per iteration, the classifier achieved an accuracy of $0.761 \pm 0.006$ 
. This represents an $\approx$2\% increase in comparison to those achieved by employing a randomly selected, balanced, training sample enclosing half of the total data ($6280$ objects). This indicates that the AL strategy has the potential to achieve comparable results with a third of the required objects for training. Such a feature will be paramount once we transition for a real data scenario with limited spectroscopic follow-up.

The main philosophy behind this module is to fully exploit details in the real data, which can result in good classification with a minimum training, as such, in order to classify real data it should ideally be allowed to perform  queries on the real data. We are currently processing past ZTF alert stream with available labels to simulate the active learning loop (Leoni \textit{et al.}, in prep).

\subsection{Ongoing Solar System objects work}\label{section:sso}

ZTF alert packets include some information about the distance and name to the nearest known Solar System object from the Minor Planet Center\footnote{\url{https://minorplanetcenter.net/}} (MPC) archive if it exists. We use this information to label alerts as Solar System object candidates. In addition, we define a filter to further identify candidates for Solar System objects that are not in the MPC archive. These candidates are fast-moving objects and should not be detected in the same coordinates after some time. We select alerts that have:
\begin{itemize}
    \item Total spatially coincident detection number is 1 or 2 ($ndethist$).
    \item If 2 detections, observations must be within 30 min.
    \item No stellar counterpart from the PanSTARRS-DR1 catalogue, \citep[$sgscore1<0.76$ ][]{Tachibana_2018}
    \item No PanSTARRS-DR1 counterpart within 1.5 arcseconds.
\end{itemize}
About 10$\%$ of all processed ZTF alerts between November 2019 and June 2020 are labelled as Solar System object candidates, and they are mostly located along the ecliptic plane, as shown in Figure~\ref{figure:footprint}.

\section{Conclusion}\label{section:summary}
In this paper we present \fink, an alert broker designed for the LSST alert stream. Our broker is the confluence between time-domain astronomy and big data, required to fully harness the power of LSST. 

Our broker's goal is to enable a wide variety of time-domain science. To enable this, it fulfils traditional broker tasks and goes beyond them by applying state-of-the-art technology and machine learning algorithms.

Fink is based on R\&D technology that is both robust and scalable to LSST's data volumes. We have tested our framework with up to $100,000$ incoming alerts per minute which is beyond the expected $20,000$ alerts per minutes for LSST. We are capable to process such volumes within minutes and keep an alert database for post-processing. Furthermore, our highly modular design is shown to allow efficient integration of existing and emergent tools as well as traceable evolution of the state of the broker.

The broker is currently deployed on the cloud and is processing the ZTF public live-alert stream. Between November 2019 and June 2020, \fink\ has received 25 million alerts and processed 8 million alerts from this public stream. All alerts (received and processed) are saved in the \fink\ database to enable post-processing.

In this work, we have shown that \fink\ is able to select microlensing, GRB counterparts and supernova candidates with current science modules using the ZTF public alert stream. These initial science cases showcase the performance of our cross-matching (catalogues and multi-wavelength surveys) and classification modules together with customisable filtering.

We are currently working on developing more science modules and improving our current ones. We invite the community for new contributions on new and existing science cases. Importantly, we are constructing web-interfaces and API services to enable a seamless user experience and enable automatising follow-up coordination with observational facilities and teams.

\fink\ is an evolving framework and this work reflects its status as of August 2020. As it is open sourced, its updated status can be found in our GitHub repository \footnote{\url{https://github.com/astrolabsoftware}}. All parts include comprehensive test suites and a general documentation with installation instructions (locally and in the cloud) and tutorials are available from the project website\footnote{\url{https://fink-broker.org/}}. Contributions and bug reports are encouraged.

\section*{Software packages used}

\fink\ makes extensive use of several libraries and frameworks among which projects from the Apache Software Foundation\footnote{\url{https://apache.org/}} (Apache Hadoop, Apache HBase, Apache Kafka, Apache Spark), astropy, numpy, matplotlib, pandas, pytorch, scikit-learn \citep{pandas:2010,numpy:2011,astropy:2018,astropy:2013,Hunter:2007,NEURIPS2019_9015,scikit-learn:2011}.

\section*{Data availability}
The data underlying this article was accessed from the \href{https://ztf.uw.edu/alerts/public/}{ZTF public alert stream}. Annotated alerts by \fink\ will be shared on request to the corresponding author due to its large volume. The framework used to process this data is open sourced: \href{https://github.com/astrolabsoftware/fink-broker}{main architecture} and \href{https://github.com/astrolabsoftware/fink-science}{science modules}. To reproduce supernova science verification results (Section \ref{section:SN2}), we have released the subset of \fink\ annotated data available in \href{https://doi.org/10.5281/zenodo.4036589}{Zenodo DOI:10.5281}, and its corresponding \href{https://github.com/astrolabsoftware/fink-usecases}{jupyter notebook}. 

\section*{Acknowledgements}

The authors are grateful to all supporters of the project. We acknowledge the support from the VirtualData cloud at Universit{\'e} Paris-Sud which provided the computing resources. The authors thank the anonymous referee for useful comments to improve this work. The authors thank the organisers and participants of the LSST Community Broker Workshop 2019 at Seattle for useful discussions on LSST transient sky aspects.  We thank the ALeRCE broker team \citep{Forster:2020} for useful discussions. The team thanks D. Muthukrishna and G. Narayan for access to their simulations and simulation configuration files. JP thanks the ANTARES broker team \citep{antares} and C. Stubens for stream access that allowed very early prototyping. 
This research has made use of "Aladin sky atlas" developed at CDS, Strasbourg Observatory, France.
The project received support from the Google Summer of Code 2019 and 2020. 
EEOI acknowledges financial support from CNRS-MOMENTUM 2018-2020.
EB gratefully acknowledge support from NASA grant 80NSSC19K0291. AB and SK acknowledge support from the European Structural and Investment Fund and the Czech Ministry of Education, Youth and Sports (Project CoGraDS-CZ.02.1.01/0.0/0.0/15003/0000437).

\small
\bibliographystyle{mnras}
\bibliography{ref}

\bsp	
\label{lastpage}
\end{document}